\documentclass[fleqn,usenatbib,11pt]{mnras}

\usepackage{newtxtext,newtxmath}

\usepackage[T1]{fontenc}
\usepackage{ae,aecompl}

\usepackage[normalem]{ulem}
\usepackage{amsmath}
\usepackage{graphicx}
\usepackage{dcolumn}
\usepackage{bm}
\usepackage{epsfig}
\usepackage{latexsym}
\usepackage{color,colortbl} 
\usepackage[usenames,dvipsnames]{xcolor}
\usepackage{subfigure,float}

\usepackage{multirow}
\usepackage{textcomp}
\usepackage{array}
\usepackage[normalem]{ulem}
\usepackage{hyperref}
\usepackage{ulem}
\usepackage{verbatim}
\usepackage{url}
\hypersetup{
	colorlinks=true,
	linkcolor=red,
	citecolor=blue,
} 

\usepackage{color,soul}
\usepackage{cuted}

\newcommand{\be}{\begin{equation}}
\newcommand{\ee}{\end{equation}}
\newcommand{\bea}{\begin{eqnarray}} 
\newcommand{\eea}{\end{eqnarray}}
\newcommand{\snr}{S/N}

\title[Model Independent Reconstruction of Galaxy Stellar Velocity Map]{Model Independent Reconstruction of Galaxy Stellar Velocity Map} 

\author[M. Denissenya et al.]{
	Mikhail Denissenya$^1$\thanks{Email: mikhail.denissenya@nu.edu.kz},
Eric V.~Linder$^{1,2,3}$, Sangwoo Park$^{4,5}$, Arman Shafieloo$^{4,5}$,  Satadru Bag$^{6,7} $
\\
\\
$^1$ Energetic Cosmos Laboratory, Nazarbayev University, Astana 010000, Kazakhstan\\ 
$^2$ Berkeley Center for Cosmological Physics, University of California, Berkeley, CA 94720, USA\\ 
$^3$ Lawrence Berkeley National Laboratory, Berkeley, CA 94720, USA\\ 
$^4$ Korea Astronomy and Space Science Institute, Daejeon 34055, Korea\\ 
$^5$ University of Science and Technology, Daejeon 34113, Korea\\
$^6$ Department of Physics, TUM School of Natural Sciences, Technical University of Munich,  James-Franck-Straße 1, 85748 Garching, Germany\\
$^7$ Max-Planck-Institut fur Astrophysik, Karl-Schwarzschild-Str. 1,  85748 Garching, Germany \\
} 

\pubyear{2023}

\begin{document}
\label{firstpage}
\pagerange{\pageref{firstpage}--\pageref{lastpage}}
\maketitle

\begin{abstract}
We develop a model independent, robust method for determining 
galaxy rotation velocities across a 2D array of spaxels from 
an integral field spectrograph. Simulations demonstrate the method 
is accurate down to lower spectral signal-to-noise than standard 
methods: 99\% accurate when median $S/N=4$. We apply it to MaNGA data to construct the galaxy velocity 
map and galaxy rotation curve. We also develop a  highly efficient cubic smoothing approach that is $25\times$ faster computationally and only slightly less accurate. Such model independent methods could 
be useful in studying dark matter properties without assuming a 
galaxy model. 
\end{abstract}




\section{Introduction}

Galaxy internal dynamics provide crucial insights into the gravitational field within the galaxy, especially regarding the distribution of its dark matter component. Early indications of dark matter's existence arose in several pivotal works in the early twentieth century, including \citet{Oort1932}, \citet{Zwicky1933}, and \citet{Babcock1939}. (\citet{Bertone2018} presents an extensive review of the historical development of dark matter; see also \citet{2023arXiv230906390B}.) Subsequently, the presence of dark matter in galaxies is firmly established through seminal studies of several spiral galaxies \citep{Rubin1970, Roberts1973, Rubin1980}, where it is observed that stars follow rotation curves that are relatively flat, rather than declining as expected from the observed light. Collectively, these studies lead to the conclusion that the dynamical mass, estimated from the motion of stars or galaxies, exceeds what could be expected based solely on visible light, thereby suggesting the presence of a dark matter component. 

To this day, internal dynamics remains a crucial aspect of galaxy studies, woven into the empirical scaling properties that include the Tully-Fisher relation \citep{TullyFisher}, Faber-Jackson relation \citep{FaberJackson}, and the fundamental plane \citep{Gudehus,Cole1994}, which are adhered to by nearly all galaxies. Above all, our comprehension of dark matter distribution in galaxies continues to rely predominantly on the precise measurements of line-of-sight dynamics \citep{Rubin1978, Bosma1981, vanAlbada1985,2011MNRAS.415..545T,2020MNRAS.496.1857L,2021MNRAS.503.5238K}; see \cite{Sofue2001} for a review.

On the observational front, integral field spectrographs have 
greatly added detail to galaxy dynamics, delivering a 
2D array of spaxels -- pixels across the face of the 
galaxy, each with its own spectra. Several large-scale surveys, including CALIFA (\href{https://califa.caha.es/}{https://califa.caha.es/})  \citep{2012A&A...538A...8S,Califa}, MANGA (\href{https://www.sdss.org/surveys/manga/}{https://www.sdss.org/surveys/manga/}) \citep{Manga2015,2015AJ....149...77D,2016AJ....152..197Y,2017AJ....154...86W} and SAMI (\href{http://sami-survey.org/} {http://sami-survey.org/}) \citep{SAMI,Croom2021},  have amassed extensive databases encompassing thousands of 
galaxies\footnote{Consider also the forthcoming survey Hector \citep{Hector} in this regard.} . 
These resources underscore the critical significance of accurately measuring the internal dynamics of galaxies.

Traditionally, galaxy kinematics are determined through spectral fitting \citep{Cappellari2004,Fernandes2005,Ocvirk2006,Walcher:2006hd,Koleva:2009kt,Sanchez,2017MNRAS.466..798C}, offering a wealth of information beyond the line-of-sight velocities, e.g.\ the details about stellar populations, dispersion, ages, metallicities etc.\  \citep{2012IAUS..284...42A,2017MNRAS.466..798C,2019A&A...622A.103B,2021ApJS..254...22J}. Nevertheless, the outcomes can be influenced by the assumptions embedded in the spectral modeling. Furthermore, the template fitting approach often encounters challenges in measuring line-of-sight velocities for low signal-to-noise ($S/N$) spectra, a scenario commonly encountered when dealing with low surface brightness galaxies or at the outskirt regions of galaxies. One potential alternative is to combine data from multiple spaxels, e.g. using Voronoi tessellations \citep{voronoi,Cappellari2003,2015A&A...573A..59G,2019MNRAS.489..608F,2020MNRAS.493.3081R,2021MNRAS.507.2488G}, to improve the  $S/N$. However, this approach introduces constraints on the precision of the results and their sensitivity across different radial distances.

In our previous work \citep{paper1}, we introduced an innovative, substantially model independent technique that relies on spaxel cross-correlation following iterative smoothing. This method aggregates information from all sections of the spectra and thus exhibits tremendous potential in handling extremely low  $S/N$ spectra,  $S/N$ $\sim$ 1. That initial work was restricted to spaxels along an axis of the galaxy to facilitate the derivation of 1D rotation curves through a Hamiltonian Monte Carlo-based fitting.
In this current work, we have expanded upon this approach and extended it to encompass the full 2D Integral Field Unit (IFU) data, enabling the determination of internal dynamics in 2D. This enhanced method calculates the velocity differences between spaxels by cross-correlating Doppler-shifted spectra while incorporating various crosschecks and symmetry considerations. Consequently, it provides a robust assessment of velocities and their associated uncertainties across the entire galaxy map. 

Section~\ref{sec:method} describes the method and its advantages, 
in particular success even for relatively low signal-to-noise ($S/N$) 
spectra. We test it extensively against simulations in 
Section~\ref{sec:sim}, to determine accuracy vs $S/N$ and explore variations of the iterative method. 
In Section~\ref{sec:manga}, we apply the technique to actual 
data from MaNGA (Mapping Nearby Galaxies at Apache Point 
Observatory \cite{Manga2015}, part of the Sloan Digital Sky Survey 4). 
We summarize and discuss further work in Section~\ref{sec:concl}.

\section{Method} \label{sec:method} 

Our method of constructing 2D line-of-sight velocity maps from the MaNGA IFU spectral data utilizes an algorithm of line-of-sight (LOS) velocity estimation between a pair of spectra, as described in \cite{paper1}. 
The LOS velocity difference $\Delta V$ between a pair of spectra (say $A$ and $B$) is determined by maximizing the weighted cross-correlation 
\be
r_{AB}(\Delta V)= \frac
{\displaystyle\sum_i w_i\, \Delta F^s_A(\lambda_i+\Delta \lambda_i)\,\Delta F_B(\lambda_i)}
{\sqrt{\displaystyle\sum_i w_i \left[\Delta F^s_A(\lambda_i+\Delta \lambda_i)\right]^2}\sqrt{\displaystyle\sum_i w_i \left[\Delta F_B(\lambda_i)\right]^2}}\ ,	
\ee 
where $\Delta\lambda=\lambda\Delta V/c$ is the wavelength shift and $\Delta F=F-\langle F\rangle$ is the weighted mean-subtracted flux. The shifted smoothed spectrum $F^s_A(\lambda_i+\Delta \lambda)$ is obtained by smoothing the original spectrum $F_A(\lambda_i)$, as discussed below, and shifting the wavelength to account for a velocity difference $\Delta V$. The uncertainties $\sigma_{Bi}$ in the unsmoothed spectrum $F_B(\lambda_i)$ are used to define the weights $w_i=1/\sigma_{Bi}^2$. 

For robustness and to enable crosschecks we split each spectrum into four equally sized wavelength bins, as described in \cite{paper1}. For each pair of spaxels, we identify overlapping wavelength intervals at a given wavelength shift and find optimal velocity differences in the corresponding bins. We then interchange the role of the two spaxels, i.e.\ smoothing and shifting the spectrum in spaxel $B$ and cross-correlating it with the observed spectrum in spaxel $A$, and repeat the calculation, expecting robust results to follow the  symmetry
\be
\Delta V_{AB}\approx -\Delta V_{BA}\ . 
\label{eq:sym}
\ee
Carrying out the mirror interchange for each of the four bins gives a total of eight $\Delta V$ estimates. 
 
We enforce the symmetry condition and require consistency of estimates between wavelength bins to provide important crosschecks on the robustness of the method, adapting the criteria of 
\cite{paper1} by 
introducing a scaling factor, denoted as crit $\times\, c$, 
\bea 
&&\left\lvert\Delta V_{AB_j}+V_{BA_j}\right\rvert\le  \label{eq:critc} \\ 
&&\qquad\max\left[\,(c\times0.05)\,\left\lvert\Delta V_{AB_j}-V_{BA_j}\right\rvert\,,\ c\times ~10~{\rm km/s}\,\right]\ . \nonumber
\eea 
We generally take $c=1$ as in 
\cite{paper1}, but study the 
velocity determination success rate for different values $c$ in Section~\ref{sec:manga}.
We obtain the final velocity $\Delta V$ and its uncertainty $\sigma_{\Delta V}$ by averaging the estimates passing the criteria and computing the standard deviation. 

For the iterative smoothing approach (see \citet{Shafieloo:2005nd, Shafieloo:2007cs, Shafieloo:2009hi, Aghamousa:2014uya}) 
we use smoothing length $\Delta=1.5$ and number of iterations $N_{\rm it}=10$. 
We also consider a faster cubic spline smoothing approach, using the \textsc{scipy.interpolate.splrep} routine available in \cite{scipy}, discussed in the next section.

\section{Testing vs Simulations} \label{sec:sim} 

Extensive testing of the iterative smoothing approach was 
presented in \cite{paper1}. Here, we explore some variations 
of that approach, with a focus on robustness, especially 
for low signal to noise ($S/N$) data, generalization from 
1D reconstruction along the galaxy major axis to the full 
2D integral field spectroscopic maps, and computational 
speedup. Therefore, we carry out further tests, initially 
with simulated data to assess accuracy. 

As an alternative to iterative smoothing of the spectral 
data, we investigated cubic interpolation, cubic spline smoothing, 
and Gaussian Process regression. The most successful alternative 
was cubic spline smoothing, so we will focus on that in this article. 

In smoothing cubic splines, input data points and their standard deviations are combined to construct a spline curve \citep{scipy}. The smoothing parameter $s$ ensures that the spline curve predictions do not result in the $\chi^2$ per degree of freedom exceeding the value $s$. We set $s=1$ to match the input data and the resulting spline curve.

\begin{figure}
	\centering
	\includegraphics[width=0.48\textwidth]{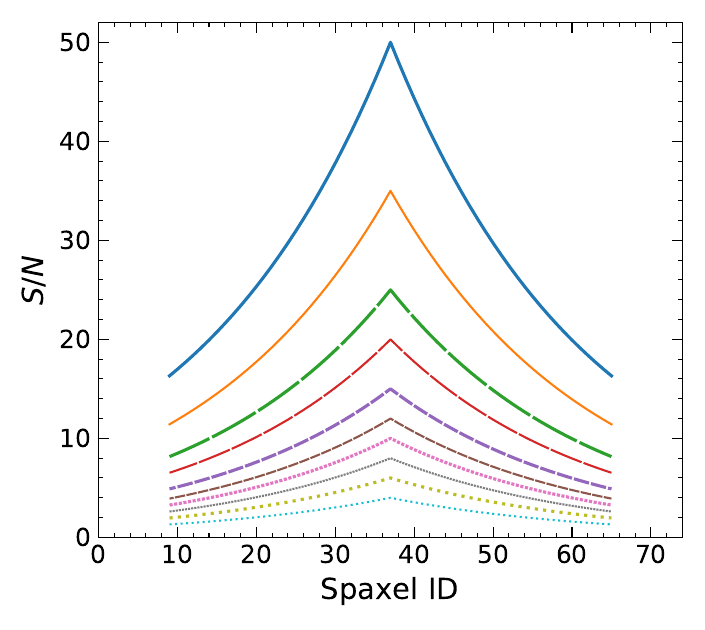}
	\caption{$S/N$ distributions used in the 1D simulations to degrade spectra in each spaxel along the major axis.  
	} 
	\label{fig:snrprof}
\end{figure}

The iterative smoothing approach was tested over a range 
of spectral  $S/N$ in \cite{paper1}, and for various spectra, but 
these were varied simultaneously. Here, we study the effect 
of  $S/N$ on velocity measurement, isolating the impact by 
fixing the spectrum for clearer quantification. We also adopt 
a more realistic model of the variation of  $S/N$ from the 
galaxy center to edge, following an exponential profile, 
\be 
S/N(r)=S/N(0)\,e^{-r/r_c}\,
\label{eq:snrf}
\ee 
where $r$ is the number of spaxels from the center. A given $S/N$ level of a spaxel at a distance $r$ from the center is obtained by varying the Gaussian noise amplitude added to the input central spaxel spectrum. We adopt $r_c=25$ for 
our simulations, a good fit for real data from 
the MaNGA 7991-12701 galaxy. 
Note that  $S/N$ for a spaxel is defined as in \cite{SNR}, 
as also used in \cite{paper1}. We simulate input rotation 
velocity curves for a range of $S/N(0)$ from 50 to 4; 
note the median $S/N\approx0.57\,S/N(0)$ for the exponential 
form, hence the median  $S/N$ ranges from 28.5 to 2.3. 
Figure~\ref{fig:snrprof} illustrates the mock measurement 
$S/N$ cases.

The simulations generate input rotation curves following 
\cite{Yoon_2021}, 
\be
V(r) = V_c \tanh\left(\frac{r}{R_t}\right)+s_{\rm out}r\,,
\label{eq:vofr}
\ee 
where $V_c=-170 $ km/s, $r$ is distance from the center in spaxels, $R_t=7.5$ in 
the same units, and $s_{\rm out}=1/R_t=0.133$. These parameters are chosen to closely match the rotation curve reconstructed from the MaNGA 7991-12701 galaxy. 

\begin{figure}
	\centering 
 	\includegraphics[width=0.48\textwidth]{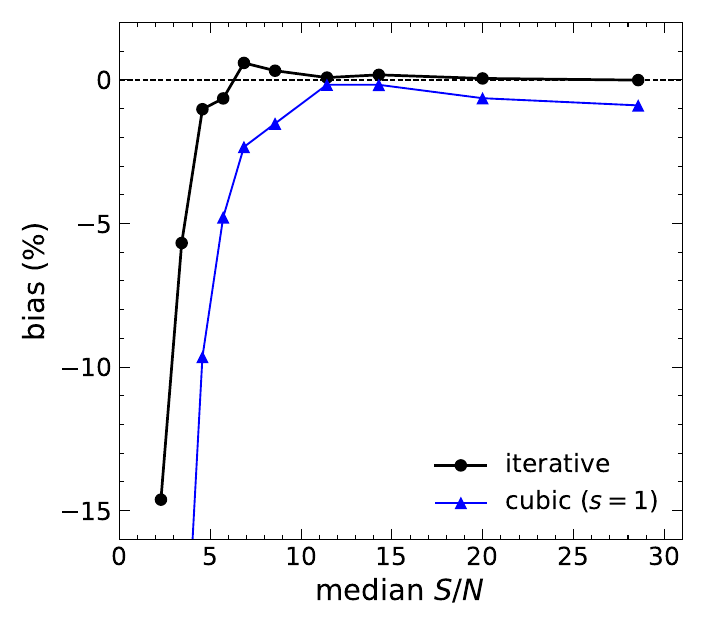}
	\caption{Reconstructed rotation curve bias relative to the simulation plotted vs median $S/N$. The black dot curve shows results for the iterative approach and the blue triangle curve is for cubic smoothing. 
The methods are accurate to $\lesssim1\%$ for median 
$S/N\ge4$ for iterative, and $\lesssim2\%$ for median $S/N\gtrsim7$ for cubic approaches. 
	} 
	\label{fig:rcbias}
\end{figure}

To quantify the results derived from the various 
fitting approaches, i.e.\ the output velocities for each 
spaxel along the 1D line of spaxels (e.g.\ major axis) 
relative to input truth, we 
explore the rms velocity offset and the accuracy, i.e.\ lack 
of bias. 
Figure~\ref{fig:rcbias} shows the accuracy in terms of the bias from the truth, 
\be
A = \frac{1}{N_{\rm fit}}\sum \frac{\Delta V_{AB}-\Delta V_{\rm true}}{\Delta V_{\rm true}}\ , 
\label{eq:bias}
\ee 
for the 10 different S/N simulations 
of Fig.~\ref{fig:snrprof}. 
The sum runs over those $N_{\rm fit}$ spaxels with  
at least two (of the four) wavelength regions 
passing the criteria in Eq.~\eqref{eq:critc} with $c=1$. 

Most informative is the accuracy, plotted in 
Figure~\ref{fig:rcbias} for both the iterative smoothing 
approach and the cubic smoothing approach. We see that 
the accuracy is within 1\% for median $S/N$$\ge4$ 
for the iterative case or 2\% for median $S/N$$\ge7$ for 
cubic smoothing (while the cubic smoothing 
approach is $\sim25\times$ faster computationally). 
Recall that the median means 
that half of the spaxels will be lower, i.e.\ the median 
$S/N=4$ case has 50\% of spaxels in $S/N=[2.3,4]$, so 
the technique is accurate down to quite low  $S/N$. 
The preference for somewhat negative bias comes from 
that the spectral feature line widths are often comparable 
to the shifts induced by the rotation velocity. Thus, in the crosscorrelation between spaxels to estimate the velocity, 
the features may overlap, preferring lower velocity shifts 
than actual for low  $S/N$ data that cannot resolve the line profile.

\begin{figure}
	\centering
 	\includegraphics[width=\columnwidth]{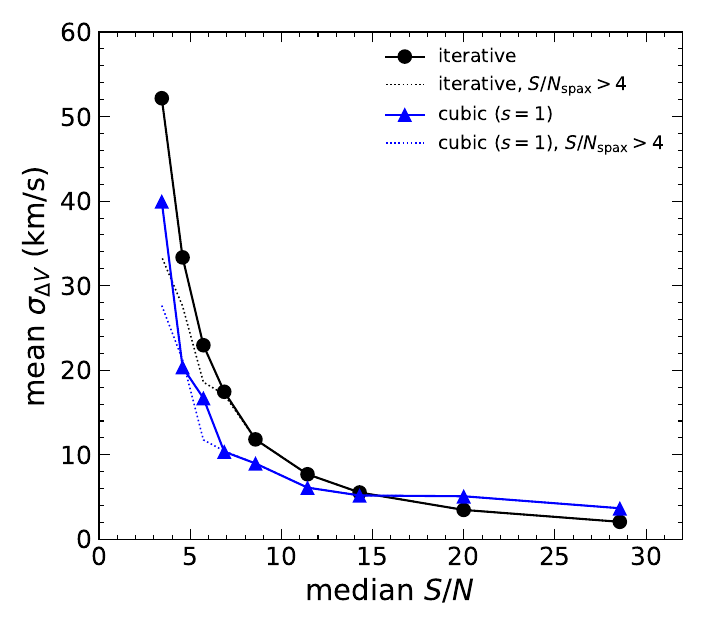}
	\caption{Rms velocity differences plotted vs median $S/N$ for the iterative (black dot curve) and cubic smoothing (blue triangle curve) approaches. Neglecting spaxels with spectra having $S/N\le4$ improves the rms by $\sim 30\%$ at low median $S/N$. 
	} 
	\label{fig:rcstdev}
\end{figure}

The reconstruction techniques also provide reasonable rms velocity uncertainties, shown in Figure~\ref{fig:rcstdev}. 
As expected, the velocity estimation has larger uncertainties 
in the low$S/N$ regions, generally at the edges of a galaxy 
where the {\it fractional\/} uncertainty will still be 
reasonable, e.g.\ $\sim10\%$. One can 
ameliorate the increase in velocity uncertainty by restricting 
to spaxels with $S/N>4$, say; we see this reduces the 
uncertainty by $\sim30-40\%$ even for median $S/N\approx3$. 
In the next section, we will also 
introduce the anchor procedure to improve robustness further.

The cubic smoothing approach is seen to be a viable 
second technique. (Note the reduced rms is just due to somewhat
fewer bins than in the iterative case passing the criteria and being included, with the good bins having less scatter.) 
Meanwhile, the iterative 
approach has been further established with tests beyond \cite{paper1}. We therefore now proceed to using 
both approaches on actual data.

\section{Application to Real Data} \label{sec:manga} 

We wish to use all the spectral data from the integral 
field spectroscopy, so we extend from the 1D major axis 
selection previously used to the full 2D hexagon of the 
MaNGA field. We focus on the MaNGA 7991-12701 galaxy whose 
major axis only was fit in \cite{paper1}.

\begin{figure*}
	\centering
 	\includegraphics[width=1.05\columnwidth]{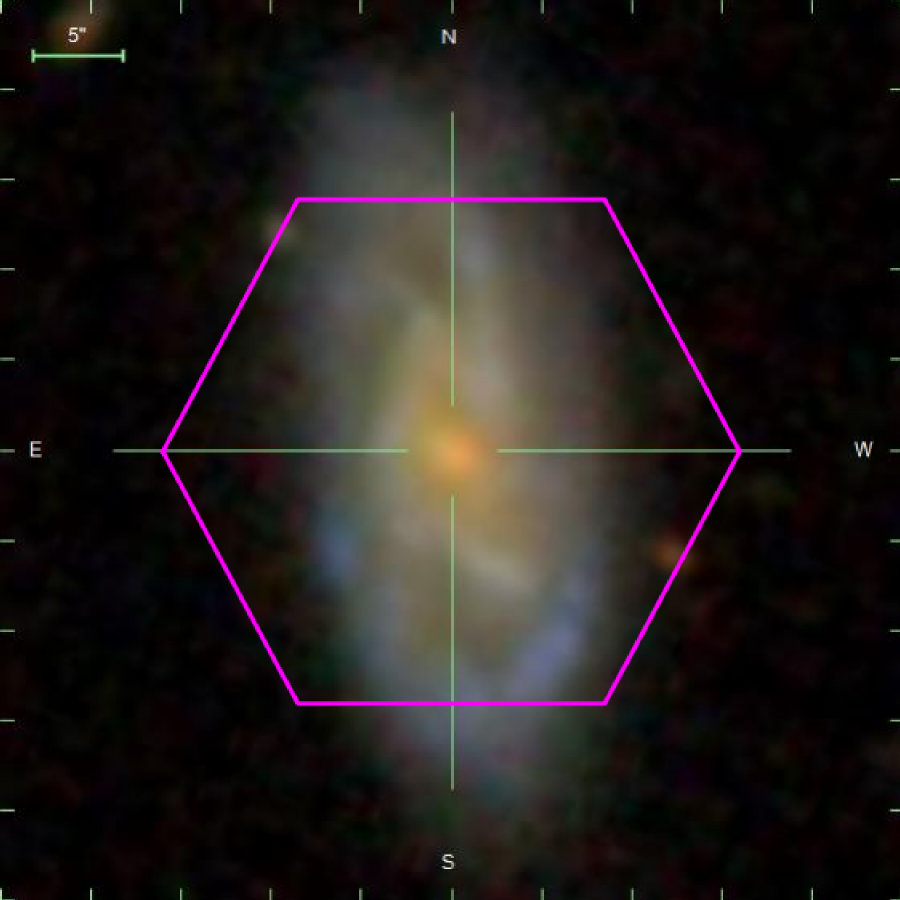}
   	\includegraphics[width=0.95\columnwidth]{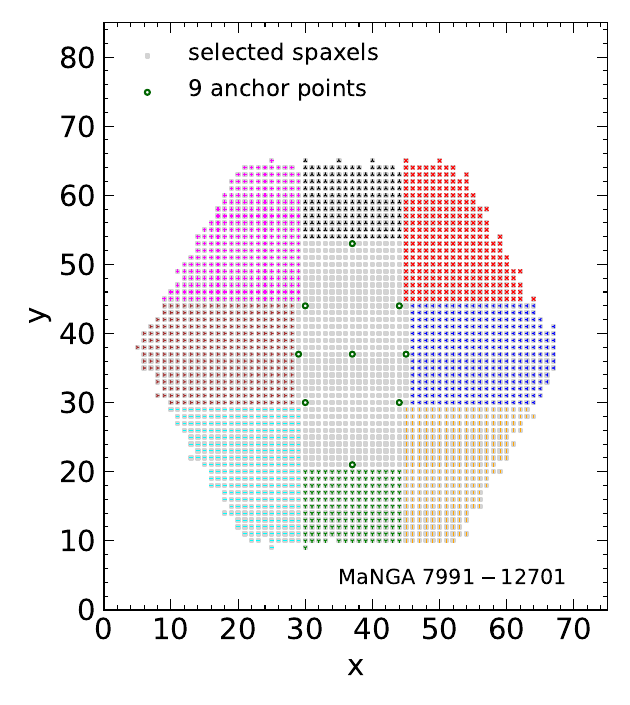}
	\caption{[Left] The optical image of the MaNGA 7991-12701 galaxy is overlaid with the hexagon of the integral field spectrograph. 
 [Right] The spaxels are shown and our nine anchor points indicated, along with the spaxel patches (colored regions) associated to each. 
	} 
	\label{fig:anchs}
\end{figure*}

MaNGA data for this galaxy comprises 2939 spaxels, but 
some of these have large gaps in wavelength coverage that 
can make the crosscorrelation technique fail or give 
spurious results. We remove spaxels with data at fewer than 3780 wavelengths, leaving 2632 spaxels. 

If one crosscorrelated every spaxel with every other one, 
this would lead to the computationally challenging task of optimizing over a million correlation functions. Instead 
we take two approaches. One is to crosscorrelate every 
spaxel with the central (usually the highest $S/N$ spaxel),  giving a $N_{\rm spaxel}$ 
rather than $N^2_{\rm spaxel}$ problem, but losing 
information on spectral variations away from the 
center. The second 
is to choose a modest set of anchor points throughout 
the galaxy, and define 
neighboring patches of spaxels ``belonging'' to an anchor point. 

Furthermore, one aspect of real data is that the spectral properties can differ 
from the center of the galaxy to its edges. A variety of 
such spectra were investigated in \cite{paper1}, to see 
the extent to which the technique there could still succeed. 
Here, the anchor method uses the spectra of 
the anchor spaxel associated with the spaxel's patch to 
estimate velocity differences, rather 
than always using the central spaxel spectrum. 

Figure~\ref{fig:anchs} shows the selection of nine 
anchor points: one is located at the galaxy center and 
eight anchors are distributed  in pairs along the galaxy 
major and minor axes and two main diagonals. 
We choose anchor spaxel distances from the center to maximize the number of their wavelength bins passing the velocity criteria Eq.~\eqref{eq:critc}, i.e.\ basically 
the anchor spectra should have good $S/N$. 
For the specific case of MaNGA 7991-12701, 
the anchors are $\pm16$ ($\pm8$) spaxels away from the center along the major (minor) axes and $\pm7$ ($\pm7$) 
spaxels away along the diagonal with a positive (negative) slope.

Thus, we show results for fits relative 
to the center (e.g.\ iterative: center), and using anchors 
(e.g.\ iterative: anchors), where 
a spaxel velocity relative to the anchor is then 
propagated to the velocity relative to the center through 
vector addition to the anchor-center velocity. 

In this article, we compare these two techniques  
for determining the velocity map, in a local approach 
(1 spaxel to 1 center 
spaxel for the center technique, 1 spaxel to 1 anchor 
spaxel to 1 center spaxel for the anchor technique). 
The anchor technique could also 
be quite useful in a global fit where the optimization is 
carried out over all spaxels simultaneously. For nine 
patches (see Figure~\ref{fig:anchs}), and optimization 
within each patch independently, this would give  
roughly $9(N_{\rm spaxel}/9)^2$ correlations. 
Hamiltonian Monte Carlo analysis is a possibility for 
such 2D simultaneous optimization, but 
such work is beyond the scope of this article. 

\begin{figure*}
	\centering
	\includegraphics[width=0.48\textwidth]{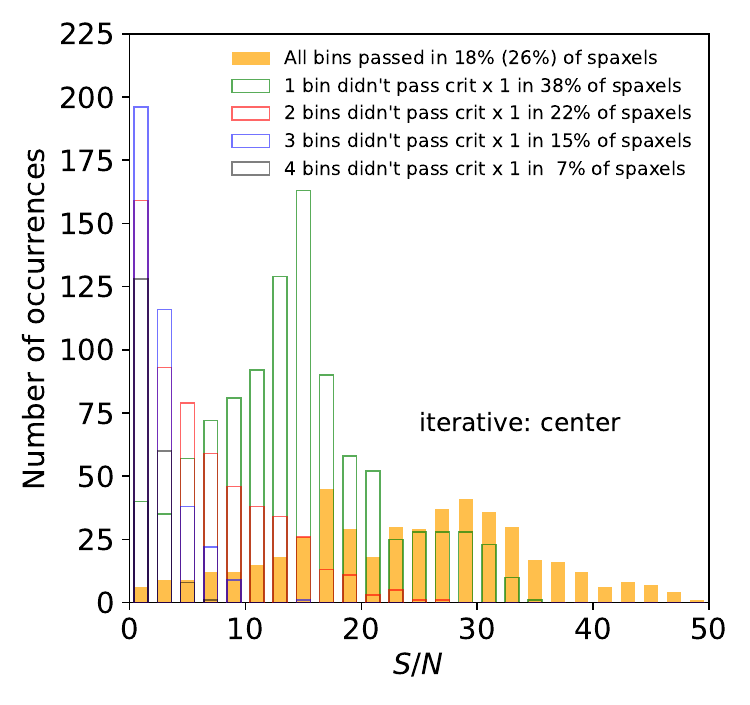} \ 
	\includegraphics[width=0.48\textwidth]{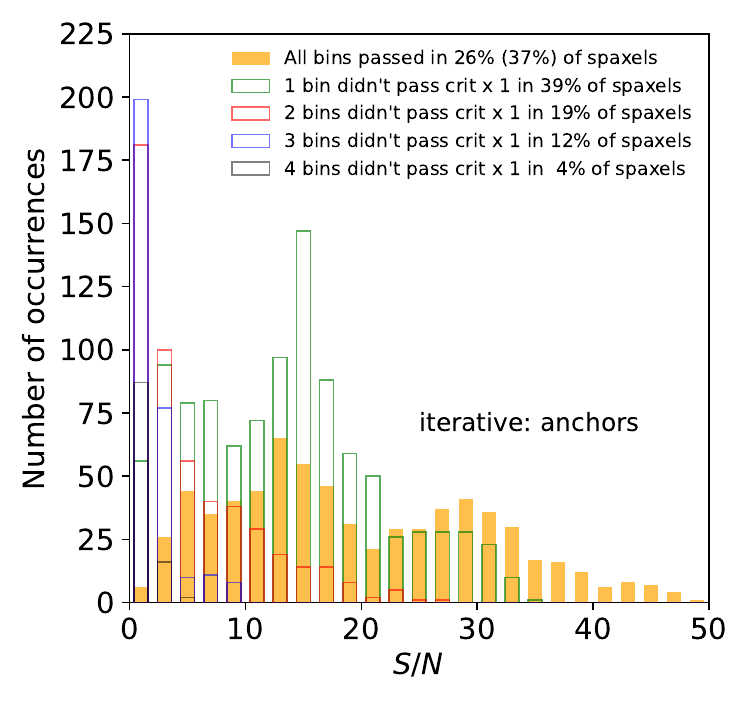} 
\caption{Distributions of signal-to-noise ratios in the MaNGA galaxy 7991–12701 spaxels, characterized by the number of bins satisfying the crit x 1 requirement on the velocity differences obtained using the iterative approach relative to the central spaxel (left), and relative to the anchor spaxels (right). 
For all bins passed case, the percent in parentheses is when considering only $S/N>4$. 
 } 
	\label{fig:gapsv2}
\end{figure*}

Carrying out the analysis of the data, we find the 
anchor technique shows improved results for the low $S/N$ 
cases. Figure~\ref{fig:gapsv2} plots the success of 
velocity fits across all wavelength bins, or a subset, 
vs $S/N$. Note that 78\% of spaxels with {\it any\/} 
$S/N$ have two or more bins 
passing the criteria Eq.~\eqref{eq:critc}, 
hence suitable for use in the 
velocity reconstruction. Looking at the range  
$S/N\ge4$, we see the number of cases not passing 
in at least two bins (i.e.\ the blue and black bars) 
rapidly diminishes (indeed $\gtrsim96\%$ pass). 
The anchor technique works even 
better, especially at low $S/N$, as seen by the increased 
height of the gold bars (all four bins pass) and diminished 
contribution of the blue and black bars (only one or zero 
bins pass).

\begin{figure*}
	\centering
	\includegraphics[width=0.48\textwidth]{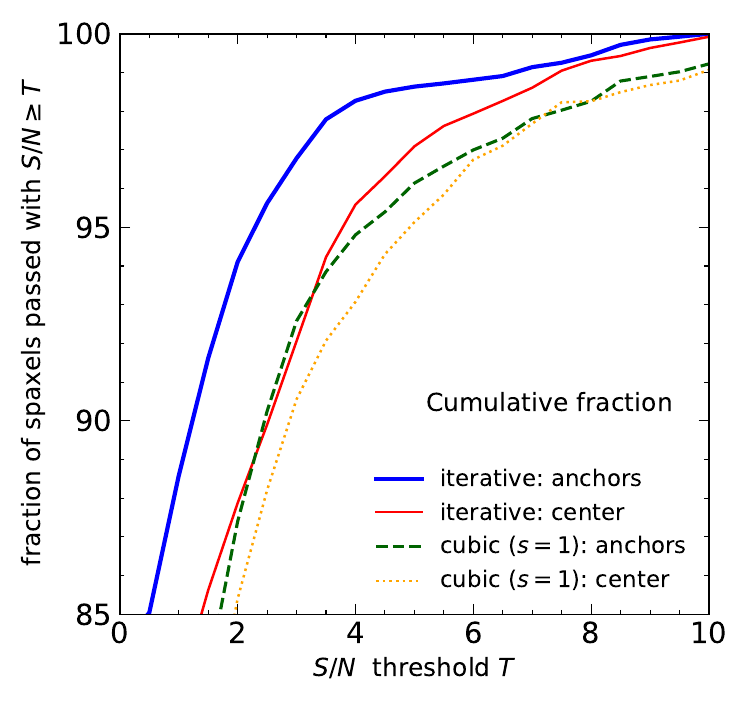} \ 
	\includegraphics[width=0.48\textwidth]{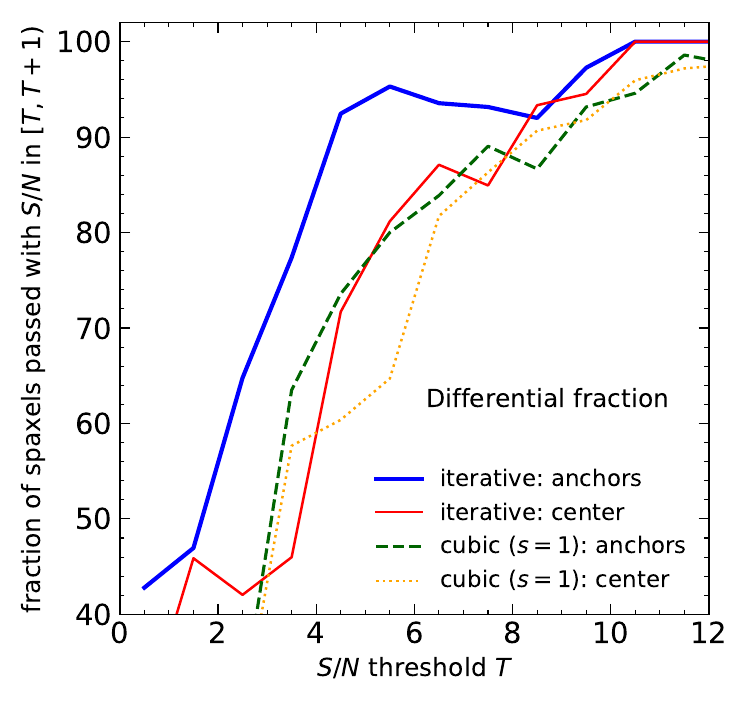}
	\caption{
 Cumulative fraction of spaxels with signal-to-noise ratios $\snr\geq T$ (left) and differential fraction of spaxels with signal-to-noise ratios $\snr$ in intervals $[T,T+1)$ 
are plotted vs the threshold $T$. 
Solid (dashed) curves show results for the iterative (cubic smoothing) approaches; thick (thin) curves show results via the anchor (central) spaxel technique. 
	} 
	\label{fig:dfxb}
\end{figure*}

Figure~\ref{fig:dfxb} summarizes the cumulative, and 
differential, fraction of spaxels that pass the required 
criteria (in at least two wavelength bins) as a function 
of $S/N$. We clearly see the advantage 
of the anchor technique, where 90\% of spaxels pass even 
if they have only $S/N=4-5$. The more computationally 
efficient cubic smoothing approach also does well. 
When considering all spaxels with $S/N$$\ge4$, the cubic 
plus anchor technique achieves 95\% success.

\begin{figure*}
	\centering
	\includegraphics[width=0.45\textwidth]{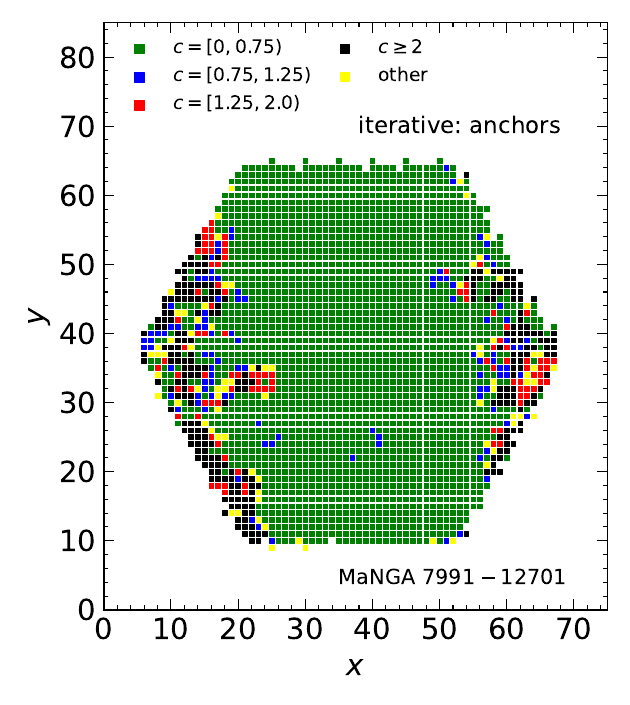}
	\includegraphics[width=0.45\textwidth]{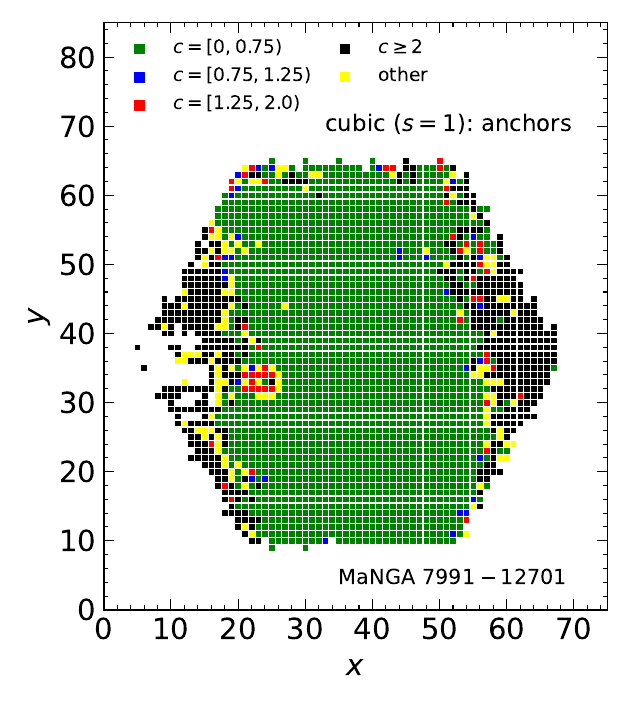}
	\caption{Success in meeting the robustness criterion for real data is shown for the 2D array of MaNGA spaxels. The color code corresponds to spaxels satisfying the criterion crit $\times\,c$ in at least two wavelength bins, for $c<0.75$, $0.75\leq c<1.25$, $1.25\leq c<2$, $c>2$. Results are given for the iterative (cubic smoothing) approach in the left (right) panel, both using 
 the anchor technique. 
Yellow points indicate spaxels where the four bins fall into four distinct $c$ ranges. 
 }
	\label{fig:cv}
\end{figure*}

We can also study how well spaxels pass the criteria, 
i.e.\ can we tighten it, or which spaxels that failed 
will pass if we loosen it. Adjusting the scaling factor 
$c$ in Eq.~\eqref{eq:critc}, we map in Figure~\ref{fig:cv} 
the velocity determination success as a function of $c$. Remarkably, 
the great majority of spaxels would pass for the anchor 
technique even if the criteria were tightened. Those 
that only pass with looser criteria ($c>1.25$ here) 
are those on the edges, far from the major axis of 
the galaxy, where the $S/N$ is lowest.

\begin{figure*}
	\centering 
	\includegraphics[width=0.48\textwidth]{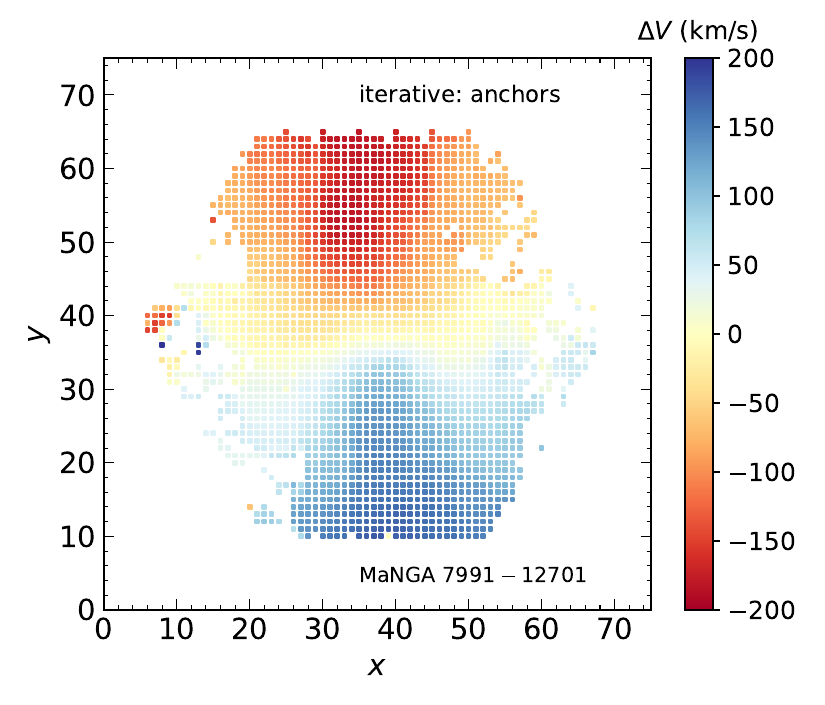}\ 
	\includegraphics[width=0.48\textwidth]{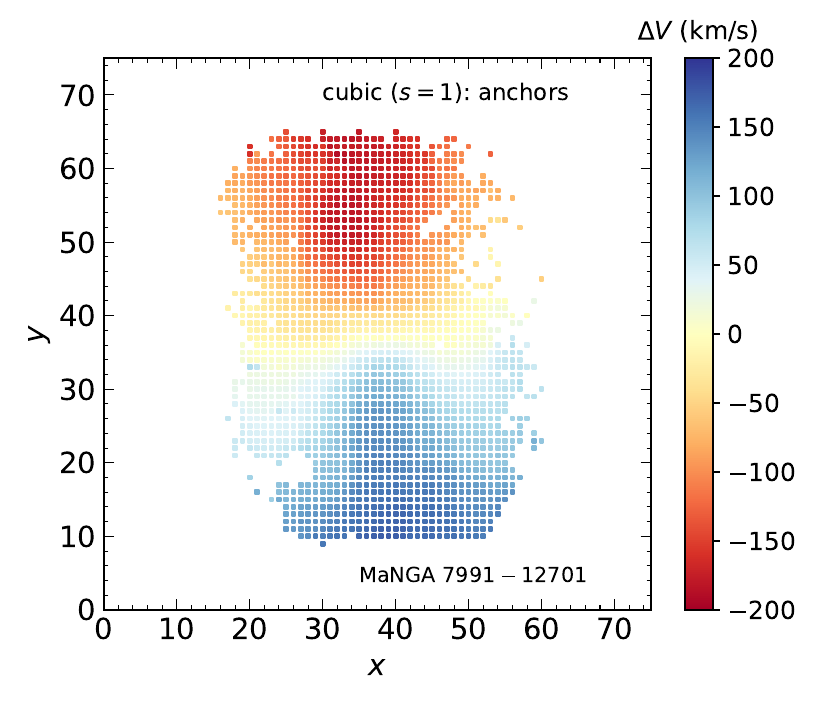}
	\caption{2D maps of the velocity $\Delta V$ relative to the center for the MaNGA 7991-12701 galaxy for the iterative (left) and cubic smoothing (right) approaches, using the anchor technique. The standard criteria $c=1$ in at least two wavelength bins are used. 
 }
 
 \label{fig:vmaps2D}
\end{figure*}

Figure~\ref{fig:vmaps2D} presents the 2D velocity map 
we reconstruct (using our fiducial criteria $c=1$) 
from the MaNGA 7991-12701 galaxy integral field 
spectroscopy data. Note that the full region with 
appreciable galaxy light (i.e.\ nonnegligible $S/N$) 
is fit; only the right and left edges where the $S/N$ 
is very low and does not pass the robustness criteria. 
Iterative and cubic smoothing techniques give 
consistent results, though iterative succeeds to 
lower  $S/N$.

\begin{figure}
	\centering
 	\includegraphics[width=\columnwidth]{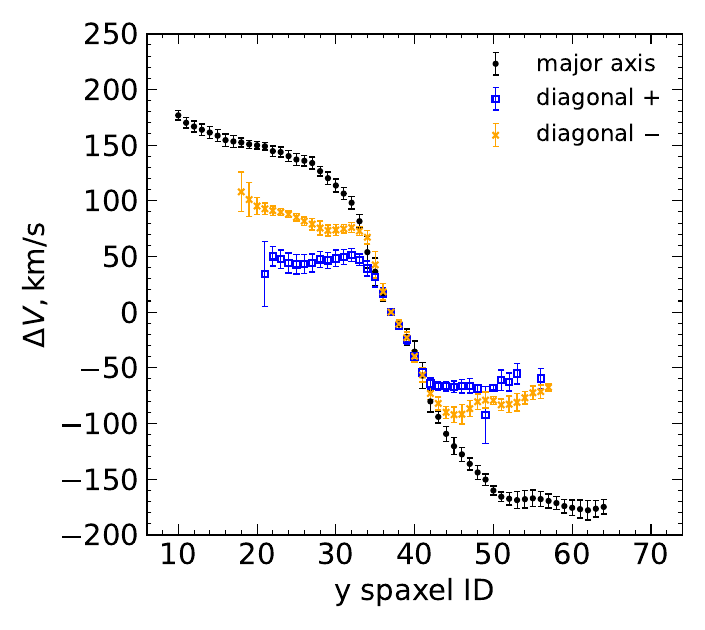}
	\caption{Reconstructed rotation curves using the iterative anchor approach along the major axis, and main diagonals with positive (negative) slope.  
	} 
	\label{fig:rcs}
\end{figure}

The 2D velocity map contains considerable useful information 
but it is often condensed to a 1D galaxy rotation curve. 
One can do this by assuming a particular model profile 
and fitting to the data, however we wish to pursue a 
model independent approach. Here we therefore present 
for illustration three slices through the 2D map, along the galaxy major 
axis and through the diagonal anchor 
points. 
Figure~\ref{fig:rcs} presents these 
results. The rotation curves along the main diagonals extend to different distances due to the galaxy inclination angle, therefore they have shorter tails compared to the rotation curve found along the major axis aligned in the $y$-direction.

\section{Conclusions} \label{sec:concl} 

Integral field spectroscopy provides rich information 
on galaxy structure, important for the understanding of 
the dark matter distribution. We develop two 
approaches for using the full 2D array of spaxels 
across a galaxy to reconstruct the velocity map without 
assuming a profile model. 

The approaches of iterative smoothing and cubic smoothing 
applied between two spaxels are successful at finding 
the relative velocity, even at low $S/N$ where standard 
techniques such as Penalized Pixel-Fitting (pPXF) \citep{Cappellari2004} may have difficulty. For 90\% of spaxels having as low as $S/N\in[4,5]$, 
for example, the fits  pass our velocity estimation 
robustness criteria. 
Tests vs simulations show the velocity reconstruction bias is 
less than 1\% down to median $S/N=4$ (i.e.\ where half of the 
spaxels have $S/N<4$). The iterative approach is more 
accurate than the cubic smoothing approach, but the latter 
is some $25\times$ faster computationally. 

The use of anchor spaxels offers several advantages. 
Robustness is greatly improved: $\sim 90\%$ of fit spaxels 
pass the criteria vs $\sim60\%$ without anchoring, and 
$\sim4\times$ more spaxels pass in all wavelengths bins, 
for $S/N\in[4,5]$. Accuracy of the overall velocity map, 
i.e.\ recovery of input in simulations, improves. For actual 
data with heterogeneity of spectra across the galaxy, 
a spaxel is likely to have a spectrum more similar to its 
neighbor anchor spaxel than to the galaxy center spaxel. 
Finally, for eventual global fits of the velocity map, 
anchors and patches could reduce dimensionality of the 
optimization and ameliorate computational burden. 

Our 2D velocity maps are smooth and well behaved, 
extending further off the major axis than the equivalent 
MaNGA-Marvin maps, and to lower $S/N$. Similarly, when 
projected to 1D galaxy rotation curves, the results are 
smoother (without any parametric model profile) and 
with generally smaller uncertainties. 
Future work includes the implementation of efficient global 
optimization for the maps, and further enhancement of 
the anchor technique.

\section*{Acknowledgments}

 We are grateful to Nazarbayev University Research 
 Computing for providing computational resources 
 for this work. This work was supported in part 
 by the Energetic Cosmos Laboratory. 
 EL is supported in part by the U.S.\ Department of Energy, Office of Science, Office of High Energy Physics, under contract no.\ DE-AC02-05CH11231. 
AS would like to acknowledge the support by National Research Foundation of Korea NRF-2021M3F7A1082053, and the support of the Korea Institute for Advanced Study (KIAS) grant funded by the government of Korea. SB acknowledges the funding provided by the Alexander von Humboldt Foundation.




\bibliographystyle{mnras}
\bibliography{gidrc}

\bsp	
\label{lastpage}
\end{document}